\documentclass{article}

\usepackage[utf8]{inputenc}
\usepackage[T1]{fontenc}
\usepackage[english]{babel}
\usepackage{textcomp}
\usepackage{amsmath,amssymb}
\usepackage{lmodern}
\usepackage{graphicx}
\usepackage[dvipsnames,svgnames]{xcolor}
\usepackage{microtype}
\usepackage{hyperref} \hypersetup{pdfstartview=XYZ}

\usepackage{hyperref}

\usepackage{caption}
\usepackage{multirow}
\usepackage{supertabular}

\usepackage{tikz}
\usetikzlibrary{patterns}

\usepackage{authblk}

\title{Atomic population kinetics in the context of XUV/X-ray free electron laser generating warm dense matter and strongly coupled plasmas}

\author{Y.J. Aouad\footnote{PhD in Plasma Physics from Pierre and Marie Curie University, Laboratory LULI. Email: aouad852000@yahoo.fr}}
\affil{\footnotesize Theoretical Physics, Atomic Physics in Plasmas, France}

\author{F.B. Rosmej}
\affil{Sorbonne Universit\'es, Pierre et Marie Curie UPMC, UMR7605, LULI, case128, F-75252 Paris Cedex 05, France}
\affil[2]{LULI, \'Ecole Polytechnique, CEA, CNRS, Physique Atomique dans les Plasmas Denses, Route de Saclay, F-91128 Palaiseau, France.}

\author{V.S. Lisitsa}
\author{A.V. Demura}
\affil{National Research Center "Kurchatov Institute", TPI, Kurchatov Square 1, Moscow 123182, Russia}

\begin{document}

\maketitle

\begin{abstract}

In the present paper we propose the use of desnity matrix formalism for the calculation of the atomic spectroscopic properties of autoionizing states in dense plasmas, i.e. warm dense matter and strongly coupled plasmas. This is motivated by the importance of taking into account the effect of the electric microfield generated by the plasma charge constituents in the calculation of the atomic populations of autoionizing states. For autoionizing states, electron densities justifying the statistical approach exceed solid density bringing serious doubts to the statistical approach even in dense plasmas.  This leads also to the change of the spectral line shapes in the frame of the standard methods. We discuss the importance of the present analysis in view of the radiation emission originating from dense plasmas created by the interaction of the X-ray Free Electron Lasers (XFEL’s) with solid density matter.  
  
\end{abstract}

\tableofcontents
\bigskip


\section{Introduction}

\subsection*{XFEL’s lasers: characteristics and advantages}
\hspace{2mm} X-ray Free Electron Lasers (XFEL’s) (installations: LCLS 2011, XFEL 2011, SACLA XFEL 2011) offer for the first time, to the community of dense plasmas, the opportunity to create matter under extreme conditions of pressure and density \cite{Rosmej1}\cite{Youcef1}. The XFEL’s are characterized by intensity of order of $10^{+16}$ W/cm$^{2}$, short pulse length of order of $10-100$ fs, photon energy of order of $1-20$ keV and high repetition frequency of order of some $10$ Hz. These properties allow the creation of new dynamical regimes of matter never achieved in laboratory so far. For example, when the XFEL’s laser interacts with a solid-state matter, the absorption of the radiation energy of the laser is homogeneous and proceeds into the volume of the solid, as the critical density of the XFEL’s is larger than the solid density. Then the absorption of the energy occurs on a short time scale when the matter is in its solid-state density allowing isochoric heating of the solid due to the characteristic time of the hydrodynamic expansion which is larger than the pulse duration of the XFEL’s.

\subsection*{XFEL’s interaction with a crystal: creation of exotic regimes of matter}
\hspace{2mm} In the interaction of the XFEL’s laser with a crystal, exotic matter regimes could be created \cite{Rosmej1}-\cite{Galtier1}. Indeed, the internal shells of most crystal atoms are photoionized in super strong laser field, and thus, for instance, the so-called hollow crystal exotic regime could be formed. The hollow atomic states are de-excited by Auger effect (autoinization), and the photo- and Auger electrons destroy the crystal arrangement of the atoms and lead to the formation of so-called Warm Dense Matter (WDM) \cite{Lee1}, which evolutes to the Dense Strongly Coupled Plasma regime (DSCP) \cite{Ichimaru1}. These regimes of matter are of interest in different areas of research like for instance in planetary science, astrophysics and all plasma-production devices where the plasma generation starts from cold dense matter (e.g., laser solid matter interaction, heavy ion beam driven plasmas, exploding wires, and pinch plasmas). However, these regimes are difficult for experimental study and still not well understood theoretically. From the physical point of view it is essential that the ionic coupling parameters $\Gamma_{ii}$ (the ratio of the Coulomb potential energy of two ions over the mean kinetic energy of ions) is larger than 1 in these regimes. Thus as the kinetic energy of ions is not larger than their Coulomb electrostatic interaction, no small parameter could be identified and the models based on perturbation theory are not applicable. Physically and experimentally theses regimes are transient and characterized by very short time intervals, so that their identification by a streak camera (time resolution $> 0.5$ ps) is hardly possible.

\subsection*{X-ray emission originating from autoionizing atomic states and hollow ion configurations: time-resolved emission}
\hspace{2mm} Nevertheless the spectroscopic methods give some way for the study of dense plasma regimes. And, indeed, it is based on the properties of the XFEL’s itself, namely the creation of multi-excited states and hollow ion configurations by direct photoionization of the K and L atomic shells, as the energy per photon that constitutes the XFEL’s allows the ionization of the internal shells of atoms (ex: K and L) \cite{Rosmej2}-\cite{Rosmej5}. These configurations are created principally in the dense plasma regime during the laser heating of matter. The autoionizing configurations (and their corresponding spectral emission) are characterized by a very short lifetimes of about $1-10$ fs due to high autoionization rates ($\Gamma \approx 10^{+13} - 10^{+16}$ s$^{-1}$). The short lifetime leads likewise to an intrinsic time resolution for the corresponding dielectronic satellite emission \cite{Rosmej1}\cite{Youcef1}\cite{Rosmej6}. The high intensity of the XFEL’s laser causes high number of photoionization events of the internal atomic shells and high population of multi-excited states and hollow ion configurations. That is why the development of new spectroscopic methods based on the multi-excited and hollow ions atomic configurations is enormously advantageous for the study of the coupled plasma regimes. 

\section{Density matrix approach for atomic kinetics in plasmas}
\hspace{2mm} In quantum mechanics, interacting systems are governed together by one Schrödinger equation. This is the case of the emitting ion inside correlated plasma, where the influence of the plasma on the ion is important for the description of the modification of atomic structure of the ion and at the same time as atomic level population dynamics \cite{Youcef1}. The latter case could be treated within a theory of open quantum systems  \cite{Meystre1}, i.e., emitting ion, in interaction with its reservoir environment (R), i.e., plasma or surrounding ions, electrons, radiation field and the quantized electromagnetic field of the vacuum. The time evolution of the total closed system (emitting ion + plasma + vacuum), from the initial time $t’ = 0$ to a given time $t$, is described by a unitary operator $\hat{U}(t,0)$. For a total Hamiltonian $\hat{H}$ which do not depend on time, one writes: 
\\
\begin{equation}\label{Evolution_operator}
\hat{U}(t,0) = \exp\left({-\frac{i}{\hbar} \hat{H} t}\right)
\end{equation}
\\
In equation Eq.\ref{Evolution_operator}, $\hbar$ is the reduced Planck constant and $\hat{H}$ is the sum of three parts: non perturbed emitting ion Hamiltonian $\hat{H}_0$, Hamiltonian of the environment (plasma + vacuum) $\hat{H}_{R}$ and the Hamiltonian $\hat{H}_{int}$ of the interaction between the above two parts. The form of the time evolution operator from equation Eq.\ref{Evolution_operator} represents a one parameter Lie group U(1) generated by the total Hamiltonian $\hat{H}$. This leads to symmetry between time evolution (of the system in question) to the future and to the past (both evolutions are realized by the same operator) and then no dissipative phenomena could be studied by such a transformation. This is correct for the total system, as this one represents a closed system. But in the case of the emitting ion (open quantum system), the realization of the time evolution must be changed into a non-unitary operator in order to take into account the real sense of the interaction of the emitting ion with its environment plasma and electromagnetic field of the vacuum. 
\\
\\
Indeed, if one is interested only on the behavior of the emitting ion, the formalism of the density matrix \cite{Fano1} is more convenient and allows deducing one part from the whole system. In this formalism, the density matrix $\rho_{tot}(t)$ of the total system at time $t$ is given by propagating the information from time $t’ = 0$ by using the unitary evolution operator $\hat{U}(t,0)$ and getting: 
\\
\begin{equation}\label{rho_Evolution_operator}
\rho_{tot}(t) = \hat{U}(t,0) \rho_{tot}(0) \hat{U}^{\dag}(t,0)
\end{equation}
\\
The Hamiltonian equation of motion that governs the total density matrix $\rho_{tot}(t)$ is given by the so-called Von Neumann equation:  
\\
\begin{equation}\label{rho_Von_Neumann}
\frac{\partial \rho_{tot}(t)}{\partial t} = -\frac{i}{\hbar} \left[\hat{H}, \rho_{tot}(t) \right] 
\end{equation}
\\
The reduced dynamics of the emitting ion density matrix $\rho$, taking into account the influence of its reservoir environment is given by taking a trace over the reservoir degrees of freedom (partial trace), making the reduced dynamics non-unitary as expected:
\\
\begin{equation}\label{reduced_rho}
\rho= Tr_{R} \left(\rho_{tot}(t) \right) 
\end{equation}
\\
Contrary to the equation Eq.\ref{rho_Von_Neumann}, the Markovian equation of evolution of the reduced density matrix Eq.\ref{reduced_rho} is realised by a one-parameter dynamical non-unitary semi-group \cite{Lindblad1} generated by a certain super-operator $\hat{L}$:
\\
\begin{equation}\label{rho_Liouvillian}
\frac{\partial \rho}{\partial t} = \hat{L} \left[ \rho \right] 
\end{equation}
\\
The limitation on a semi-group is caused by the fact that dissipative processes always tend to equilibrium or other stationary states. Therefore inverse propagations are not feasible and thus corresponding propagators (backwards in time) does not exist, at least not practically computable ones. In equation Eq.\ref{rho_Liouvillian}, the super-operator $\hat{L}$ split in two parts:
\\
\begin{equation}\label{split_Liouvillian}
\hat{L} \left[ \rho \right] = \hat{L}_{unitary} \left[ \rho \right]  + \hat{L}_{dissipative} \left[ \rho \right] 
\end{equation}
\\
In equation Eq.\ref{split_Liouvillian}, $\hat{L}_{unitary} $ represents the unitary part of the equation of motion induced by a Hermitian operator representing a renormalized Hamiltonian due to the interaction with the environment, i.e., modification of the atomic structure and electric field mixing of atomic states. $\hat{L}_{dissipative}$ provides the dissipative evolution parts, i.e., relaxation of atomic populations due to the collisions with the electrons and due to the spontaneous emission etc. The Markov approximation is introduced by considering that the relaxation time of the environment to its equilibrium state (after a perturbation coming from the emitting ion) is very small in comparison to the time of relaxation decay of the emitting ion, no-memory reservoir approximation. The determination of the form of the super-operator $\hat{L}$ is not unique and depends on the studied physical system.  
\\
\\
The direct application of the equation Eq.\ref{split_Liouvillian} to the emitting ion, interacting with its plasma and vacuum environments, is difficult. The difficulty is multiple; the first one is related to the complexity of the density matrix formalism as the system of equation concerns diagonal elements and non-diagonal ones, which increase considerably the rank of the total system of equations \cite{Youcef1}. The second one concerns the theoretical expressions of the super-operator $\hat{L}$ and particularly the dissipative parts. For some atomic transition processes an explicit expression exists and is based on a certain approximations. But other processes are introduced heuristically and are deduced only by analogy to those that are calculated more or less explicitly \cite{Rautian1} -\cite{Anufrienko2}. A certain theoretical rigorous work is needed here.


\section{Explicit expression of the atomic density matrix kinetic equation in plasmas}
\hspace{2mm} To write an explicit expression of the operator $\hat{L}$ for numerical calculations \cite{Youcef1}, the effect of the plasma on the atomic kinetics is introduced and concerns both, the non dissipative Hamiltonian  mixing of the atomic levels by the quasi-static ion electric microfield and the dissipative time evolution part induced by the multiple atomic processes in the plasma. The later part is often introduced in the frame of the collisional radiative kinetics model where in the density matrix case the difference is given by the relaxation of the non-diagonal density matrix elements also called coherences. The explicit form of the time evolution equation of the density matrix $\rho$ reads:
\\
\begin{equation}\label{ATM_tev_dm}
\frac{\partial \rho}{\partial t} = -\frac{i}{\hbar} \left[\hat{H}_0+\hat{V}(\vec{E}), \rho \right] +\hat{\Lambda}\left[\rho\right]
\end{equation}
\\ 
where $\hbar$ is the reduced Planck constant and the first term of the right hand side of Eq.\ref{ATM_tev_dm} includes  $\hat{H}_{0}$ the unperturbed Hamiltonian of the emitting ion, $\hat{V}$ the interaction of the emitting ion with the quasi-static electric microfield $\vec{E}$ of the surrounding ions in the dipole approximation. More precisely, the use of the quasi-static approximation is justified by the fact that the relaxation time of the density matrix elements is smaller than the inverse plasma frequency associated to the ions. Denoting the dipole atomic moment of the emitting ion $\hat{\vec{d}}$, the interaction potential energy $\hat{V}$ is given by: 
\\
\begin{equation}\label{V_op}
\hat{V}(\vec{E})=-\hat{\vec{d}} \cdot \vec{E}
\end{equation}
\\
The dissipative part $\hat{\Lambda}\left[\rho\right]$ in equation Eq.\ref{ATM_tev_dm} includes all the collisionnal and radiative relaxation processes that populate and depopulate the elements of the density matrix, i.e., spontaneous emission rate $A$ and autoionization rate $\Gamma$, collisionnal relaxation between levels that are mixed by the electric microfield $\vec{E}$ of ions and so on. This constitute an approximate expression of $\hat{L}_{dissipative}$ in equation Eq.\ref{split_Liouvillian}. It is to note that the solution density matrix of Eq.\ref{ATM_tev_dm} is field dependent: 
\\
\begin{equation}\label{rho_E}
\rho \equiv \rho(\vec{E})
\end{equation}
\\
The final result must to be averaged over an electric microfield distribution function $Q(\vec{E})$:
\\
\begin{equation}\label{rho_average}
<\rho>=\int d\vec{E} \ Q(\vec{E}) \rho(\vec{E})
\end{equation}


\section{Atomic population interpretation of the diagonal density matrix elements}
\hspace{2mm} The diagonal density matrix elements are related to the atomic populations which in the presence of the plasma electric microfield mixing is subject to the choice of the representation of the projection basis of the density matrix operator into matrix elements. This is related to the choice of the unperturbed atomic basis: unperturbed atomic quantum states $\mid \gamma  J M > $ (states of total kinetic momentum $J$, its quantized projection $M$ and additional quantum numbers $\gamma$ depending on the choice of the coupling scheme) or unperturbed atomic energy levels $\mid \gamma  J > $. In the first case we relate the diagonal density matrix elements to populations of atomic quantum states $N_{\gamma  J M}$ and int the second case to populations of atomic energy levels $N_{\gamma  J}$. Both populations are related to each other by conservation principle:
\\
\begin{equation}\label{Stat_sum_pop}
N_{\gamma  J}=\sum_{-J \leqslant M \leqslant J}N_{\gamma  J M}
\end{equation}

\subsection{Labeling of atomic quantum states in the presence of plasma electric microfield: M as a good quantum number}
\hspace{2mm} In the presence of the plasma electric microfield $\vec{E}$, the interaction with the emiting ion in the dipole approximation reads \cite{Cowan1}:
\\
\begin{equation}\label{dip_ion_int}
\hat{V}(\vec{E})=-\vec{E} \cdot \textbf{\emph{P}}^{\textbf{(1)}}
\end{equation}
\\
where, $\textbf{\emph{P}}^{\textbf{(1)}}$ is the dipole operator of the emiting ion written as an irreducible tensor operator of rank 1. Because of the odd parity of the dipole operator it leads that its matrix elements are non-zero only between states of opposite parity and therefore to belong to two different configurations. By representing the electric field in the z-direction axis and using the Wigner-Eckart theorem, the non-diagonal matrix elements of $\hat{V}(\vec{E})$ are given by:
\\
\begin{multline}\label{ME_H_elec}
\left\langle \gamma J M \mid \hat{V}(\vec{E}) \mid\gamma ' J' M' \right\rangle = -2 \ E \ (-1)^{J-M} \left(\begin{array}{clcr}
J & 1 & J'\\
-M & 0 & M'  \end{array}\right) \times \\
< \gamma J \mid\mid \textbf{\emph{P}}^{\textbf{(1)}} \mid\mid \gamma ' J' > 
\end{multline}
\\
where, $< \gamma J \mid\mid \textbf{\emph{P}}^{\textbf{(1)}} \mid\mid \gamma ' J' >$ is the reduced matrix elements. The properties of the 3-j symbol lead to the condition of equality of $M$ and $M'$ implying that $M$ remains a good quantum number. In the $\text{LS}$ coupling choice of representing the unperturbed quantum states, the reduced matrix elements $< \ \mid \ \mid \ >$ which is $M$ independent is simplified due to the commutation of the spin operator with the electric dipole operator and is given by:
\\ 
\begin{multline}\label{Red_ME_Dipole}
< \gamma J \mid\mid \textbf{\emph{P}}^{\textbf{(1)}} \mid\mid \gamma ' J' > = \delta_{S,S'} (-1)^{L+S+J'+1} [J,J']^{1/2} \left\lbrace \begin{array}{clcr}
L & S & J\\
J' & 1 & L'  \end{array}\right\rbrace \times \\
< \gamma J S \mid\mid \textbf{\emph{P}}^{\textbf{(1)}} \mid\mid \gamma ' J' S' >
\end{multline} 
\\
where the last term of Eq.\ref{Red_ME_Dipole} contains the transition matrix elements:
\\
\begin{equation}\label{P_nlnl}
\textbf{\emph{P}}_{nl,n'l'}^{(1)} = (-1)^{l} [l,l']^{1/2} \left(\begin{array}{clcr}
l & 1 & l'\\
0 & 0 & 0  \end{array}\right) \int_{0}^{\infty} P_{nl}(r)P_{n'l'}(r) \ r\  dr
\end{equation}
\\
where $P_{nl}(r)$ is the radial function of the one-electron spin-orbitales having the principal quantum number $n$ and the orbital quantum number $l$.
\\
\\
Following the property of the projection of the plasma electric microfield dependent interaction $\hat{V}(\vec{E})$, namely non-zero matrix elements only between unperturbed atomic states having the same projection magnetic quantum number $M$, one has to construct atomic density matrix elements in the basis $\mid \gamma  J M > $ and to relate it to the quantum state atomic population $N_{\gamma  J M}$ and then to construct populations  $N_{\gamma  J}$ of atomic energy levels.


\subsection{Construction of field dependent quantum states} 
\hspace{2mm} When taking into account the interaction of the emitting ion with the plasma electric microfield, the quantum states of the system are designated by their quantized projection of their total kinetic momentum $M$ as was seen in the previous subsection. In a pur $\text{LS}$ coupling sheme, the unperturbed states $\mid \beta L S J M >$ are eigenstates of the unperturbed Hamiltonian $\hat{H}_0$ corresponding to the unperturbed energie eigenvalues $E_{\beta L S J M }$ (where corresponding to this scheme we put $\gamma \equiv \beta L S$):
\\
\begin{equation}\label{H0_energies}
\hat{H}_0 \mid \beta L S J M > = E_{\beta L S J M } \mid \beta L S J M >
\end{equation}
\\
In the presence of the interaction, the field dependent quantum states $\mid \alpha M(E) >$ are eigenstates of the total Hamiltonian corresponding to the field dependent eigenvalues $E_{\alpha M(E) }$:
\\
\begin{equation}\label{HE_energies}
(\hat{H}_0 + \hat{V}(\vec{E})) \mid \alpha M(E) > = E_{\alpha M(E) } \mid \alpha M(E) >
\end{equation}
\\
Now we expand the field dependent quantum states in the unperturbed basis $\mid \beta L S J M >$:
\\
\begin{equation}\label{Coeff_C}
\mid \alpha M(E) > = \sum_{\beta, L, S, J} C_{\beta L S J}^{\alpha M}(E) \mid \beta L S J M >
\end{equation}
\\
where, the coefficients $C_{\beta L S J}^{\alpha M}(E)$  are called mixing coefficients of the quantum state $\mid \alpha M(E) >$.


\subsection{Construction of field dependent density matrix elements}
\hspace{2mm} As seen in the previous two subsections, the general idea is to construct atomic populations from the diagonal density matrix elements and to relate one to the other. As a first step, the unperturbed quantum states of the unperturbed atomic system $\mid \gamma J M >$ are used for the projection of the density operator $\rho$ into matrix elements representation: 
\\
\begin{equation}\label{rho_matrix_elements}
\rho^{\gamma J,\  \gamma' J'}_{M,\  M'} = < \gamma J M \mid \rho \mid \gamma ' J' M' > 
\end{equation}
\\
From Eq.\ref{rho_matrix_elements}, the time evolution of the matrix elements of the density matrix are obtained by the projection of Eq.\ref{ATM_tev_dm} \cite{Youcef1}:
\\
\begin{multline}\label{rho_tev_matrix_elements}
\frac{\partial \rho^{\gamma J,\  \gamma' J'}_{M,\  M'}}{\partial t} = -\frac{i}{\hbar} (E_{\gamma J M }-E_{\gamma' J' M' } )\rho^{\gamma J,\  \gamma' J'}_{M,\  M'}- \\
\frac{i}{\hbar} \sum_{\gamma'',J'', M''}\left(\hat{V}(\vec{E})^{\gamma J,\  \gamma'' J''}_{M,\  M''} \times \rho^{\gamma'' J'',\  \gamma' J'}_{M'',\  M'}-\rho^{\gamma J,\  \gamma'' J''}_{M,\  M''} \times \hat{V}(\vec{E})^{\gamma'' J'',\  \gamma' J'}_{M'',\  M'}\right)+ \\
\hat{\Lambda}\left[\rho\right]^{\gamma J,\  \gamma' J'}_{M,\  M'}
\end{multline}
\\ 
where:
\\
\begin{equation}\label{V_matrix_elements}
\hat{V}(\vec{E})^{\gamma J,\  \gamma' J'}_{M,\  M'} = < \gamma J M \mid \hat{V}(\vec{E}) \mid \gamma ' J' M' > 
\end{equation}
\\
and:
\\
\begin{equation}\label{lambda_matrix_elements}
\hat{\Lambda}\left[\rho\right]^{\gamma J,\  \gamma' J'}_{M,\  M'} = < \gamma J M \mid \hat{\Lambda}\left[\rho\right] \mid \gamma ' J' M' > 
\end{equation}
\\
The relaxation term writes:
\\
\begin{multline}\label{rho_relaxation_elements}
\hat{\Lambda}\left[\rho\right]^{\gamma J,\  \gamma' J'}_{M,\  M'} =-\frac{1}{2}\left(\sum_{\gamma'',J'', M''}W^{\gamma J,\  \gamma'' J''}_{M,\  M''} +\sum_{\gamma'',J'', M''}W^{\gamma' J',\  \gamma'' J''}_{M',\  M''} \right) \rho^{\gamma J,\  \gamma' J'}_{M,\  M'}+\\
\delta_{\gamma,\gamma'} \delta_{J,J'} \delta_{M,M'} \sum_{\gamma'',J'', M''}W^{\gamma'' J'',\  \gamma J}_{M'',\  M} \times \rho^{\gamma'' J'',\  \gamma'' J''}_{M'',\  M''}
\end{multline}
\\
where the coefficients $W^{\gamma'' J'',\  \gamma J}_{M'',\  M} $ are the standard collisional radiative rates of atomic processes in plasmas corresponding to the transition from the atomic state $\mid \gamma'' J'' M'' > $ to $\mid \gamma  J M > $.
\\
\\
In equations Eq.\ref{rho_tev_matrix_elements}-\ref{rho_relaxation_elements} where the density matrix projection is given in the field independent quantum states $\mid \gamma L S J M >$, the master equation takes into account the non-dissipative Hamiltonian part of the interaction of the emitting ion with the plasma electric microfield and hence populations and coherences, are subject to both, the non-dissipative Hamiltonian part and the dissipative part (collisional-radiative like).
\\
\\
It is to note that if we project the density operator into matrix elements using the field dependent quantum states basis $\mid \alpha M(E) >$ then the density matrix master equation will not contain the non-disspative Hamiltonian evolution part related to the unperturbed Hamiltonian and the interaction of the emitting ion with the plasma electric microfield as both of these parts are already included in the definition of atomic quantum states. In this case the projection of the density operator $\rho$ into matrix elements is given by: 
\\
\begin{equation}\label{rho_matrix_elements_field}
\rho_{\alpha M(E),\  \alpha' M'(E)} = < \alpha M(E) \mid \rho \mid \alpha ' M'(E) > 
\end{equation}
\\
And the projection of the master equation Eq.\ref{ATM_tev_dm} leads to: 
\\
\begin{equation}\label{rho_tev_matrix_elements_field}
\frac{\partial \rho_{\alpha M(E),\  \alpha' M'(E)}}{\partial t} = < \alpha M(E) \mid\hat{\Lambda}\left[\rho\right]\mid \alpha ' M'(E) > 
\end{equation}
\\   
The last equation Eq.\ref{rho_tev_matrix_elements_field} is more difficult to be established as it involves field dependent wave functions Eq.\ref{Coeff_C} and the evaluation of the collisional-radiative processes in plasma in a field dependent manner. This is a still open question in the community of atomic physics in dense plasmas. The equations Eq.\ref{rho_tev_matrix_elements}-\ref{rho_relaxation_elements} are more convenient for numerical calculations in order to take the influence of the plasma electric microfield on atomic populations.


\section{Atomic density matrix of autoionizing states}
\hspace{2mm} When many levels system is introduced for numerical calculations, the rank of the system Eq.\ref{rho_tev_matrix_elements} become very large and even more complicated than the standard collisionnal-radiative model that is already prohibitive for numerical calculations \cite{Youcef1}\cite{Youcef3}\cite{Youcef4}\cite{Rosmej7}. The difficulty arises firstly from the fact that atomic quantum states $\mid \gamma J M >$ are used instead of atomic energy levels $\mid \gamma J >$ and secondly from the nature of the density matrix formalism that introduce coherences (non-diagonal matrix elements) in addition to populations (diagonal matrix elements) \cite{Youcef1}. The situation is more complicated when the autoionizing atomic configurations are considered in the calculation. These kinds of configurations contain a huge number of atomic levels of different $\gamma J$ terms as they have an open atomic sub-shells and various coupling schemes of the momentums of the bound electrons have to be implemented. These autoionizing configurations have a high autoionization rate (of some $10^{+14}$ to $10^{+16}$ s$^{-1}$ for the configuration $K^{1}L^{8}M^{Y}$) and then their statistics is very far from Boltzmann distribution \cite{Youcef1}. In order to present the spectra of the autoionizing configurations, that is important for understanding of  the high-density regimes (WDM and DSCP), it is necessary to calculate their atomic populations $N_{\gamma J}$ from the density matrix model. Due to the complexity of both: the system Eq.\ref{rho_tev_matrix_elements} and the autoionizing configurations, the only possibility that allows to introduce in the radiative-collisionnal model the effect of mixing, due to the non-diagonal elements of the density matrix, is to simplify the system of equations Eq.\ref{rho_tev_matrix_elements} by rewriting it in terms of atomic energy levels $\mid \gamma J >$ and then to find an analytical expression that express the non-diagonal matrix elements in terms of diagonal ones \cite{Youcef1} in order to arrive to a system of kinetic equations on populations that include the effect of plasma electric microfield mixing of levels.  


\section{Conclusion}
\hspace{2mm} In this paper we have presented the general formulation of the density matrix applied to atomic kinetics in dense plasmas. The model is implemented in the atomic quantum states basis $\mid \gamma J M >$ and a discussion on the autionizing atomic states is given. This method could allow studying the influence of the correlations between ions on the atomic populations in a strongly coupled plasmas (DSCP) and in the warm dense matter (WDM). 
      


\end{document}